\documentclass{an}
\usepackage{graphicx}
\usepackage{times}
\usepackage{fancyhdr}
\def\degree{\nobreak\ifmmode{^\circ}\else{$^\circ$}\fi}

\newcommand{\ltsimeq}{\raisebox{-0.6ex}{$\,\stackrel        
        {\raisebox{-.2ex}{$\textstyle <$}}{\sim}\,$}}

\newcommand{\gtsimeq}{\raisebox{-0.6ex}{$\,\stackrel
        {\raisebox{-.2ex}{$\textstyle >$}}{\sim}\,$}}

\def\grtsim{\mathrel{\hbox{\rlap{\hbox{\lower2pt\hbox{$\sim$}}}\raise2pt\hbox{$>$}}}}
\def\lesssim{\mathrel{\hbox{\rlap{\hbox{\lower2pt\hbox{$\sim$}}}\raise2pt\hbox{$>$}}}}

\begin{document}

\title{Most supermassive black hole growth is obscured by dust}

\author {Alejo Mart\'\i nez-Sansigre$^{1}$, 
Steve Rawlings$^{1}$, Mark Lacy$^{2}$, Dario Fadda$^{2}$,
Francine R. Marleau$^{2}$, Chris Simpson$^{3}$, Chris J. Willott$^{4}$ and 
Matt J. Jarvis$^{1}$}
\institute{Astrophysics, Department of Physics, University of
Oxford, Keble Road, Oxford OX1 3RH, UK \and
Spitzer Science Center, California Institute of Technology, MS220-6, 1200 E. California Boulevard, Pasadena, CA 91125, USA \and
Department of Physics, University of Durham, South Road, Durham DH1 3LE, UK \and
Herzberg Institute of Astrophysics, National Research Council, 5071 West Saanich Rd, Victoria, B.C. V9E 2E7, Canada}

\date{Received; accepted; published online}

\abstract{We present an alternative method to X-ray surveys for
hunting down the high-redshift type-2 quasar population, using Spitzer
and VLA data on the Spitzer First Look Survey. By demanding objects to
be bright at 24 $\mu$m but faint at 3.6 $\mu$m, and combining this
with a radio criterion, we find 21 type-2 radio-quiet quasar
candidates at the epoch at which the quasar activity peaked.  Optical
spectroscopy with the WHT confirmed 10 of these objects to be type-2s
with $1.4 \leq z \leq 4.2$ while the rest are blank.  There is no
evidence for contamination in our sample, and we postulate that our
11 blank-spectrum candidates are obscured by kpc-scale dust as opposed to dust from
a torus around the accretion disk. By carefully modelling our
selection criteria, we conclude that, at high redshift, 50-80\% of the
supermassive black hole growth is obscured by dust.  }

\correspondence{ams@astro.ox.ac.uk}

\maketitle

\section{Introduction}

Periods of exponential growth of supermassive black holes are observed
as quasars in the distant Universe. These quasars, however, are only
seen to outshine their host galaxy if the geometry of the dusty torus
surrounding the accretion disk is at a favourable angle
(Antonucci 1993). These are known as type-1 quasars, which
show a spatially-unresolved ``blue bump'' in the UV-optical wavebands,
as well as broad emission lines. The activity of such type-1 quasars
is known to increase with lookback time, and seems to have peaked
around $z \sim 2$ (Wolf et al. 2003). However, the
understanding of the obscured quasar population (known as type-2
quasars) is far less complete, and it is not known how much growth
occurs in obscured regions.

One of the problems in understanding the amount of obscured growth is
that there are several discrepancies in the ratio of type-2 to type-1
quasars, depending on the wavelength at which these are
observed. 

\begin{itemize} 
\item Unified schemes, where type-1 and type-2 objects are the same
intrinsically but viewed at a different orientation with respect to
the torus, predict type-2 to type-1 ratios $\sim 1$, for a
half-opening angle of $\sim40\degree$. Studies of radio-loud AGN are
consistent with this picture (Willott et al. 2000).
\item The summed emission from quasars  generates the cosmic X-ray background, 
and modelling of this background is consistent with a population of type-2 
quasars that outnumber the type-1s by $\sim $3:1 (Worsley et al. 2004). 
\item Deep X-ray surveys are sensitive to intrinsically faint AGN down
to the Seyfert regime and have found an obscured population but this
population is not large enough to account for the unresolved X-ray
background (Zheng et al. 2004; Barger et al. 2005).
\end{itemize}

Possible explanations for these discrepancies include a number of
Compton thick quasars (e.g. Alexander et al. 2005), which are
invisible even to deep X-ray surveys, and another type of optical
obscuration unrelated to the torus. With objects such as IRAS F10214
in mind (Rowan-Robinson et al. 1991), and following the success of
mid-infrared selection at lower redshift (Lacy et al. 2004; 2005a), we devised an alternative method for looking for
radio-quiet type-2 quasars at the peak of the quasar activity ($z \sim
2$). In order to be able to compare this population to the type-1
quasars, we used well defined selection criteria.

\section{Selection Criteria}

Our first step in choosing a method of hunting for the elusive high
redshift type-2 population is to consider the effects of dust on the
spectral energy distribution (SED) of a type-1 quasar. Figure 1 shows
the X-ray to far-infrared SED for two quasars, one with $A_{\rm V} =
0$ (a type-1) and another with $A_{\rm V} = 10$ (a moderate extinction
for a type-2).  We see that the effect of dust is negligible at
(observed) far- to mid-infrared, but suddenly becomes very severe in
the mid- to near-infrared. The far- to mid- infrared SED of a type-1
and type-2 are thus almost identical, while the type-2 quasar SED will
be dominated by the host galaxy at near-infrared to UV
wavelenghts. With this in mind, we used the 1.4 GHz, 24-$\mu$m and
3.6-$\mu$m data from the Spitzer First Look Survey (Condon et
al. 2003; Marleau et al. 2004; Lacy et al. 2005b; Fadda et al. in
prep.) and the following selection criteria (Mart\'\i nez-Sansigre et
al. 2005):

\begin{enumerate}
\item $S_{24 ~ \mu \rm m} > 300~\mu$Jy  
\item $S_{3.6 \mu \rm m} \leq 45~\mu$Jy
\item $350 ~\mu \rm Jy \leq S_{1.4 \rm~GHz} \leq 2$ mJy
\end{enumerate}

The mid-infrared criterion $S_{24 \mu \rm m} > 300~\mu$Jy was chosen
to select quasars below the ``break'' in the luminosity function at $z
\sim 2$. The peak of the quasar activity occurred at this redshift and
by targeting the quasars around the ``break'' we would be sensitive to
the dominant part of the population.  At $z = 2$, the break in the
quasar luminosity function, $L^{*}_{quasar}$, corresponds to $M_{\rm
B} = -25.7$ (Croom et al. 2004, with Pure Luminosity Evolution), so
assuming a typical quasar SED (Rowan-Robinson 1995) our 24-${\mu
\rm m}$ selection will select quasars just below the break in the
luminosity function ($\gtsimeq 0.2 ~ L^{*}_{quasar}$) at $z = 2$, and
more luminous quasars at higher redshifts.

\begin{figure}
\resizebox{\hsize}{!}
{\includegraphics[]{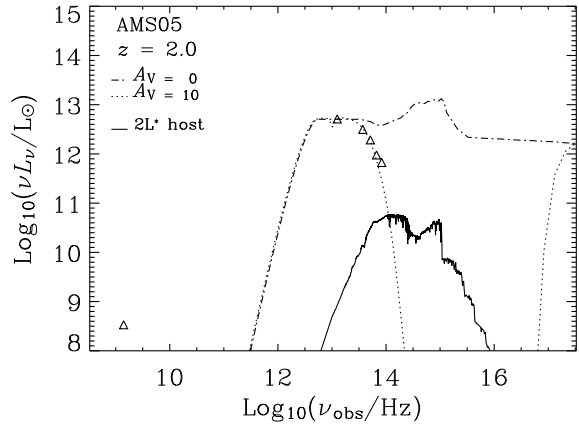}}
\caption{The spectral energy distribution (SED) of AMS05, a $z=2$
  type-2 quasar (fit by eye). The data points are (left to right)
  observed 1.4 GHz (bottom left) and 24, 8, 5.8, 4.5 and 3.6 $\mu$m
  from the Spitzer Space Telescope.  Type-1 SED (dashed line) is from
  Rowan-Robinson (1995). We have removed the far-infrared `bump' as we
  have no information on it. To obtain the type-2 SED (solid line) we
  applied the dust model from Pei (1992). The host galaxy is $2L^{*}$
  from Bruzual \& Charlot (2003). We can see that for an extinction of
  $A_{\rm V}> 10$ the 24 $\mu$m flux density will be identical for a
  type-1 and a type-2, while the 3.6 $\mu$m flux density will be
  dominated by the host galaxy, allowing us to apply a `3.6 $\mu$m-$z$
  relation' to obtain rough photometric redshifts. }
\label{figlabel1}
\end{figure}

Quasars are normally considered as being type-2 if they have an
$A_{\rm V} \gtsimeq 5$ (Simpson et al. 1999) which will make
their observed near-infrared emission much fainter than that of
type-1s. The 3.6-$\mu \rm m$ criterion was therefore chosen to remove
naked (type-1) quasars as well as lower redshift ($z \ltsimeq 1.4$)
type-2s.  Dust extinction ensures that type-2 quasars are much fainter
than type-1 quasars, even for a moderate $A_{V}$.  Indeed, the $S_{3.6
\mu \rm m}$ emission (rest-frame 1-2 $\mu$m) is likely to be dominated
by starlight for $A_{V} \geq 10$, and since light at 3.6-$\mu$m is
dominated by the old stellar population, there will be an $S_{3.6 \mu
\rm m}-z$ correlation, analogous to the $K-z$ relation for radio
galaxies (Willott et al. 2003). Assuming that host galaxies for
$z = 2$ radio-quiet quasars have a luminosity of $2L^{*}_{gal}$ 
(Kukula et al. 2001), we adapt the $K-z$ relation to $2L^{*}_{gal}$
hosts at 3.6-$\mu$m.  This criterion corresponds to a limiting
`photometric redshift' $z_{\rm phot} \gtsimeq 1.4$. This was chosen to
target $z \sim 2$ type-2 quasars, while allowing for scatter in the
photometric redshift estimation and filtering out type-1 quasars and
low-redshift contaminants like radio galaxies.  The infrared selection
criteria are plotted schematically in Figure 2.

The radio selection criteria is added to ensure that the candidates
are radio-quiet quasars rather than starburst galaxies. The 3.6-$\mu$m
- 24-$\mu$m ``colour'' we are demanding can be achieved by lower
redshift ($z \leq 1$) ULIRGs.  To avoid such starburst contaminants,
we chose a lower limit on $S_{1.4{\rm GHz}}$ well above the level
reached by high-redshift starburst galaxies (Chapman et al. 2005) 
as well as an upper limit to filter out the
radio-loud objects, whose extended jets might complicate
interpretation. We do not demand a detection at 3.6-$\mu$m, allowing
for very high-redshift objects. In addition, the FLS radio positions
(Condon et al. 2003), accurate to $\sim 0.5$ arcsec, were
better for spectrosocopic follow-up than the 24-$\mu$m positions.

\begin{figure}
\resizebox{\hsize}{!}
{\includegraphics[angle = 90]{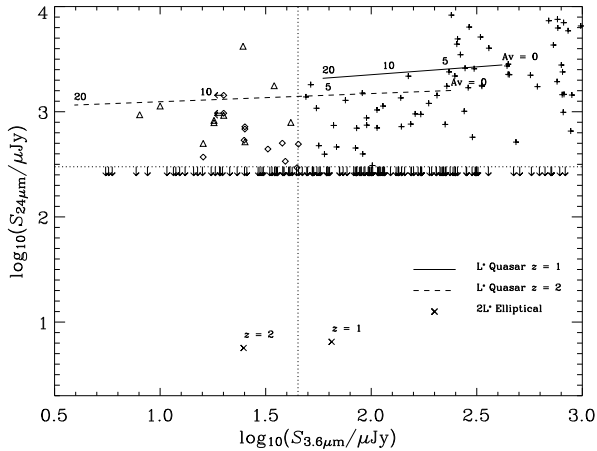}}
\caption{ 
(from Mart{\'{\i}}nez-Sansigre et al. 2005) The sources plotted
  all obey the radio criterion, so they are all AGN.  Crosses and
  upper limits are sources that do not make the infrared
  criteria. Triangles are sources with spectroscopic redshifts,
  diamonds are sources with blank spectra or no lines. The track for a
  $z=1$ quasar (solid line) shows that any quasar bright enough to
  make the 24-$\mu$m criterion (upper half of the figure) will be too
  bright to make the 3.6-$\mu$m criterion if it has an extinction
  between $A_{\rm V} = 0$ and $A_{\rm V} \sim 25$. If the extinction
  is larger ($A_{\rm V} > 25$) then the 3.6 $\mu$m emission will be
  dominated by the host galaxy (the cross at $z=1$) which is still too
  bright to make our cut.  A quasar at $z=2$ (dashed line) bright
  enough to make the 24-$\mu$m criterion will also be too bright
  unless it has an $A_{\rm V} \grtsim 6$.  
  Any quasar with $A_{\rm V} \grtsim 10$ will, in addition, 
  have its 3.6 $\mu$m light dominated 
  by the host galaxy. This allows us to apply a rough photometric
  redshift estimate, based on the $K-z$ relation. 
}
\label{figlabel2}
\end{figure}

\section{Observational Results}

These selection criteria led to 21 candidate type-2 radio-quiet
quasars in the 3.8 square degrees of the Spitzer First Look Region
that have coverage with all three wavebands. To obtain optical
spectra, the 21 candidates were looked at for $\sim 30$ minutes each
with the ISIS instrument at the WHT.  This yielded 10 spectroscopic
redshifts in the range $1.4 \leq z \leq 4.2$, 10 blank spectra and one
with faint red continuum. The spectrum of the highest redshift object
is shown in Figure 3. All the objects with redshifts showed narrow
emission lines and no type-1 quasar or lower redshift starbursts were
found to contaminate the sample (Mart\'\i nez-Sansigre et
al. 2005; Mart\'\i nez-Sansigre et al. in prep.). In the remainder of this
paper we will refer to the candidates with spectroscopic redshifts as
narrow-lined objects, and the rest as blank objects.

Eight of the objects showed narrow Lyman-$\alpha$ with equivalent 
width $>100 ~ \rm nm$, and several objects showed higher-excitation lines. 
The object with faint red continuum is probably a type-2 with $z < 1.7$ so 
that  Lyman-$\alpha$ is not visible in the optical, and has no other 
lines bright enough to be detected. There is therefore no evidence for 
contamination in our sample.

\section{Modelling the Population}

To place our results in context, we modelled the expected number of
type-1 quasars with $z \geq 2$ which would meet our 24 $\mu$m and 1.4
GHz selection criteria. We used a type-1 luminosity function (LF) from
the COMBO-17 survey (Wolf et al. 2003) and a model SED
(Rowan-Robinson 1995) to convert between rest-frame B-band
luminosity and observed 24 $\mu$m flux density. To model the radio
selection, we used an optical-to-radio correlation with scatter 
(Cirasuolo et al. 2003).  From this modelling, we would expect
$4.3^{+2.2}_{-1.1}$ type-1 quasars at $z \geq 2$ matching our
mid-infrared and radio criteria. This number is small mainly due to
the radio cut: we are looking at the radio-bright end of the
radio-quiet quasar population. Without the radio criterion we would
expect $\sim 55$ type-1 quasars.

\begin{figure}
\resizebox{\hsize}{!}
{\includegraphics[angle=90]{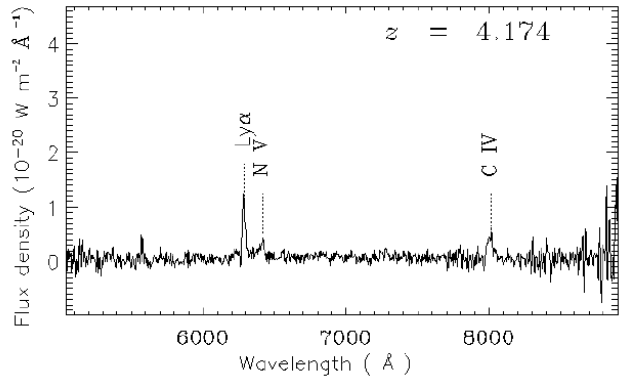}}
\caption{ (from Mart\'\i nez-Sansigre et al. 2005) The spectrum of
AMS16, the highest redshift type-2 in our sample, showing
high-excitation lines as well as Lyman-$\alpha$.}
\label{figlabel3}
\end{figure}

Using a Bayesian approach, we produced probability distributions for
the quasar fraction at $z \geq 2$, $q$ given our data and assuming 4.3
type-1 quasars (where $q$ is the fraction of type-1 quasars over the
total number of type-1 and type-2 quasars).  Two such distributions
were calculated (Figure 4): one where only the spectroscopically
confirmed type-2s with $z \geq 2$ were included (5 type-2s) and one
were the photometric redshifts were used for the objects with no
spectroscopic redshift (for a total of 11 type-2s with $z \geq 2$).

If we only take the redshift of the objects with lines in the spectra,
then the quasar fraction, $q$ is consistent with $\sim 0.5$ and the type-2
to type-1 ratio is $\sim $1:1. The probability distribution peaks at
$q \sim 0.25$ when the photometric redshifts are used to fill in, resulting 
in a type-2 to type-1 ratio $\sim $3:1. 

Figure 4 also includes the predicted $q$ for two receding-torus models
(Simpson 1998; 2005). These are consistent
with the quasar fraction from narrow-line objects at the 1$\sigma$ level, and with $q$
from both narrow-line and blank objects at the 2$\sigma$ level.

\section{Discussion}

We have found two possible values for the quasar fraction, $q$. The
spectroscopic redshifts are, on average, consistent with the
photometric redshifts, indicating that the host galaxies of the
narrow-line objects are consistent with progenitors of present-day
$2L^{*}$ elliptical galaxies. The narrow line objects are therefore
hosted by relatively dust-free (and hence transparent) galaxies so we
can see the narrow line region. The quasars in the narrow-line objects
are therefore obscured by the torus around the accretion disk. The
ratio of narrow-line type-2 to type-1 is consistent with this torus
picture. Type-1 quasars must also be hosted by relatively transparent
galaxies (i.e. ellipticals, Kukula et al. 2001), so we find that the quasar
fraction in high-redshift elliptical galaxies is consistent with $q
\sim 0.5$ as expected from unified schemes.

The blank objects, however, are almost certainly also in the redshift
range $1.4 \leq z \leq 5$. Our photometric redshifts are reliable on
average, and there is no hint of contamination by type-1s or
lower-redshift starbursts (which would show the [OII] 3727 \AA ~line
in the red end of the spectra or at least some bright continuum). We
are therefore confident that the blank objects are also high redshift
type-2s but the quasar is probably obscured by some dust on scales
large enough to obscure the narrow-line region as well as the
broad-line region. Such kpc-scale dust is characteristic of
starbursts, and we could therefore be seeing supermassive black holes
growing inside starbursting galaxies (Fabian 1999). The
orientation of the torus would therefore become irrelevant for the
blank objects, as the obscuring dust would be distributed along the
entire host galaxy.

\begin{figure}
\resizebox{\hsize}{!}
{\includegraphics[angle=90]{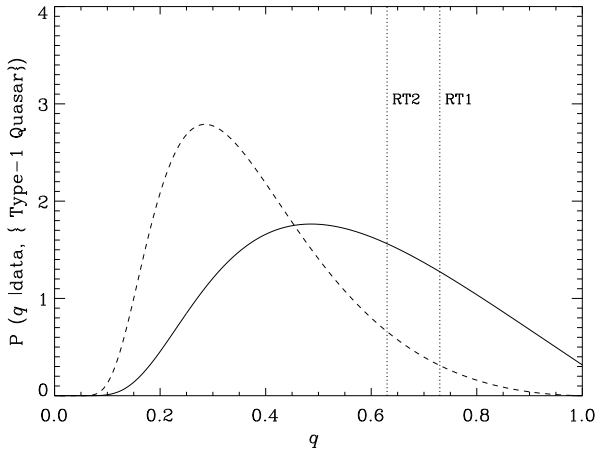}}
\caption{ (from Mart\'\i nez-Sansigre et al. 2005) The probability
distributions for the quasar fraction, $q$, given our data and
assuming 4.3 type-1 quasars at $z \geq 2$ would satisfy our 24 $\mu$m
and radio criteria. The solid line is the distribution if we only take
our type-2 quasars with spectroscopic redshift $z \geq 2$, while the
dashed line is the distribution if we add those with photometric $z
\geq 2$. The two receding-torus models labelled are RT1: Simpson
(1998) and RT2: Simpson (2005).}
\label{figlabel4}
\end{figure}

This would explain why the ratio of narrow-line (`host-obscured')
type-2s to type-1s is $\sim $1:1, while the total ratio of type-2s
(`host-obscured' and `torus-obscured') to type-1s would be $\sim $3:1,
consistent with the X-ray background
(Worsley et al. 2004). Unified schemes can only be tested in
relatively transparent galaxies, where the obscuration is due to
orientation, so the receding-torus models (Simpson 1998; 2005) must only be compared to the
curve that uses the narrow-line objects and they are
consistent. However, for the X-ray background the location of the
obscuring material is irrelevant: at high-redshift there are $\sim 3$
times more obscured quasars than unobscured.  Hence, 50-80\% of the
accretion at $2 \leq z \leq 5$ is obscured by dust . Since this is the peak of the quasar
activity and since at lower redshifts, most AGN are obscured 
(from X-ray studies: Barger et al. 2005), we can conclude that
most supermassive black hole growth is obscured by dust 
(Mart\'\i nez-Sansigre et al. 2005).

\acknowledgements

AMS, SR, ML, CS and MJJ would like to thank the organisers of the
conference: Montse, Rosa, Enrique and Jose Luis for their kind
hospitality. AMS would also like to thank the Council of the European
Union for funding, and the Ministerio de Educaci\'on y Ciencia for
financial support for this conference. SR and CS would like to thank
the UK PPARC for a Senior Research Fellowship and an Advanced
Fellowship respectively.

\end{document}